\documentclass[acmsmall,screen,nonacm]{acmart}
\usepackage{booktabs}
\usepackage{array}
\usepackage{graphicx}
\graphicspath{{figures/}}
\usepackage{multirow}
\usepackage{colortbl}
\usepackage{environ}
\usepackage{xspace}
\usepackage{pgfplots}

\acmJournal{TOSEM}
\setcopyright{none}
\newcommand{\etal}{\xspace\textit{et al.}\xspace}

\newcommand{\trackerSnapshotDate}{January 27, 2026}

\newcommand{\totalBuilds}{14,159}
\newcommand{\usableBuilds}{9,868}
\newcommand{\usedBuilds}{4,513}
\newcommand{\attemptedBuilds}{2,451}

\newcommand{\recordedExecutionHours}{1,894}
\newcommand{\hourLongExecutions}{174}

\newcommand{\trackerIssues}{51,408}        %
\newcommand{\cppIssueShare}{83.5}          %

\newcommand{\cppLogCases}{2,166}
\newcommand{\rustLogCases}{911}
\newcommand{\goLogCases}{647}
\newcommand{\jvmLogCases}{436}
\newcommand{\cLogCases}{178}
\newcommand{\pythonLogCases}{163}
\newcommand{\swiftLogCases}{12}
\newcommand{\nonCppLogCases}{2,169}

\newcommand{\ruleRecoveries}{469}
\newcommand{\autoRecoveries}{238}
\newcommand{\humanRecoveries}{231}
\newcommand{\noPatchRecoveries}{286}

\newcommand{\productiveRules}{52}
\newcommand{\singleRuleChainRecoveries}{396}
\newcommand{\multiRuleChainRecoveries}{73}
\newcommand{\multiRuleChainMedian}{2}
\newcommand{\multiRuleChainMean}{2.48}

\newcommand{\baselineGroups}{1,106}
\newcommand{\concGroups}{161}

\newcommand{\singletonGroups}{727}

\newcommand{\failRate}{69.7}

\newcommand{\concShare}{71.0}
\newcommand{\noPatchRate}{11.7}

\newcommand{\nonCppLogPct}{48.1}

\newcommand{\singletonCasePct}{16.1}

\newcommand{\topTenCasePct}{30.6} %
\newcommand{\fidelityRepaired}{52.7}
\newcommand{\fidelityNoPatch}{50.7}

\newcommand{\ageDatedCases}{1,320}
\newcommand{\ageComparedProjects}{23}
\newcommand{\ageOlderHigherProjects}{5}
\newcommand{\ageOlderLowerProjects}{10}
\newcommand{\ageTiedProjects}{8}
\newcommand{\ageMeanDifference}{-3.1}
\newcommand{\ageSignTestP}{0.302}
\newcommand{\projectComparedCount}{21}
\newcommand{\projectMinUnresolvedRate}{3.0}
\newcommand{\projectMaxUnresolvedRate}{100}
\newcommand{\ageEligibleCases}{884}

\newcommand{\autoOnlyRecoveries}{157}
\newcommand{\autoExistingOnlyRecoveries}{72}
\newcommand{\autoMixedRuleRecoveries}{9}
\newcommand{\autoDerivedRules}{29}
\newcommand{\autoCRecoveryRate}{28.8}
\newcommand{\autoJVMRecoveryRate}{23.6}
\newcommand{\autoCppRecoveryRate}{17.2}
\newcommand{\autoRustRecoveryRate}{2.3}
\newcommand{\strictControlledRecoveries}{30}

\newcommand{\controlComparableCases}{49}
\newcommand{\controlRepairOnlyCases}{24}

\newcommand{\controlBothRecoveredCases}{25}
\newcommand{\controlRepairRequiredRate}{49.0}
\newcommand{\controlRepairRequiredCILow}{35.6}
\newcommand{\controlRepairRequiredCIHigh}{62.5}
\newcommand{\controlMcNemarP}{1.2\times10^{-7}}
\newcommand{\mechanismClassifiedRecoveries}{415}
\newcommand{\mechanismEnvironmentChanges}{326}
\newcommand{\mechanismRefreshRecoveries}{89}
\newcommand{\mechanismUnclassifiedRecoveries}{54}
\newcommand{\fidelityDifference}{2.0}
\newcommand{\fidelityDifferenceCILow}{-5.3}
\newcommand{\fidelityDifferenceCIHigh}{9.3}

\newcommand{\controlEvaluatedPairs}{50}

\newcommand{\agentBuildLo}{2.3}
\newcommand{\agentBuildHi}{3.6}
\newcommand{\costPerAuto}{0.22}
\newcommand{\ruleDerivationCost}{53.07}
\newcommand{\terraCostTotal}{10.53}

\newcommand{\appliedRuleCount}{408}
\newcommand{\topReusedRuleCount}{20}
\newcommand{\topReusedRuleCreditShare}{85}
\newcommand{\crossProjectRuleCount}{26}
\newcommand{\crossProjectRuleCreditShare}{55.8}
\newcommand{\gdalMirrorRecoveries}{67}
\newcommand{\reserveNoMatchCases}{376}
\newcommand{\reserveLibreOfficeCases}{11}
\newcommand{\pairedRuleRecovered}{30}
\newcommand{\pairedQwenRecovered}{26}
\newcommand{\pairedTerraRecovered}{24}
\newcommand{\pairedRuleRate}{60.0}
\newcommand{\pairedQwenRate}{52.0}
\newcommand{\pairedTerraRate}{48.0}
\newcommand{\pairedRuleCILow}{46.2}
\newcommand{\pairedRuleCIHigh}{72.4}
\newcommand{\pairedQwenCILow}{38.5}
\newcommand{\pairedQwenCIHigh}{65.2}
\newcommand{\pairedTerraCILow}{34.8}
\newcommand{\pairedTerraCIHigh}{61.5}
\newcommand{\pairedRuleBuilds}{47}
\newcommand{\pairedQwenBuilds}{145}
\newcommand{\pairedTerraBuilds}{88}
\newcommand{\pairedRuleBuildRate}{1.57}
\newcommand{\pairedQwenBuildRate}{5.58}
\newcommand{\pairedTerraBuildRate}{3.67}

\newcommand{\pairedQwenCalls}{1,096}
\newcommand{\pairedTerraCalls}{550}
\newcommand{\pairedQwenTokens}{14.8M}
\newcommand{\pairedTerraTokens}{16.6M}
\newcommand{\pairedValidCases}{47}
\newcommand{\pairedValidRuleRecovered}{28}
\newcommand{\pairedValidQwenRecovered}{25}
\newcommand{\pairedValidTerraRecovered}{22}

\newcommand{\pairedQwenHarnessFailures}{3}
\newcommand{\pairedRuleOnlyVsQwen}{13}
\newcommand{\pairedRuleOnlyVsTerra}{13}
\newcommand{\pairedRuleUnrecovered}{20}
\newcommand{\pairedAgentOnly}{10}      %
\newcommand{\pairedAnyRecovered}{40}   %
\newcommand{\hybridQwenRecovered}{39}
\newcommand{\hybridQwenBuilds}{111}
\newcommand{\hybridQwenBuildRate}{2.85}
\newcommand{\hybridQwenTokens}{6.3M}
\newcommand{\hybridTerraRecovered}{37}
\newcommand{\hybridTerraBuilds}{83}
\newcommand{\hybridTerraBuildRate}{2.24}
\newcommand{\hybridTerraTokens}{7.2M}
\newcommand{\hybridTerraCost}{4.41}

\newcommand{\pairedMinAdjustedP}{0.789}

\newcommand{\applicationAuditCases}{124}
\newcommand{\repeatedRepairConcepts}{5}
\newcommand{\methodExhaustedCases}{30}
\newcommand{\scopeLimitedCases}{30}
\newcommand{\escalationProbeRecovered}{3}
\newcommand{\escalationEvaluatedCases}{57}
\newcommand{\rqfourQwenRecovered}{16}
\newcommand{\rqfourQwenPreviousRecovered}{14}
\newcommand{\rqfourTerraRecovered}{10}
\newcommand{\rqfourQwenRate}{28.1}
\newcommand{\rqfourQwenPreviousRate}{24.6}
\newcommand{\rqfourTerraRate}{17.5}
\newcommand{\rqfourQwenBuilds}{156}
\newcommand{\rqfourQwenPreviousBuilds}{128}
\newcommand{\rqfourTerraBuilds}{78}
\newcommand{\rqfourQwenBuildRate}{9.75}
\newcommand{\rqfourQwenPreviousBuildRate}{9.14}
\newcommand{\rqfourTerraBuildRate}{7.80}
\newcommand{\rqfourQwenCrashReproduced}{10}
\newcommand{\rqfourQwenPreviousCrashReproduced}{11}
\newcommand{\rqfourTerraCrashReproduced}{7}
\newcommand{\rqfourQwenCalls}{1,175}
\newcommand{\rqfourQwenPreviousCalls}{1,352}
\newcommand{\rqfourTerraCalls}{638}
\newcommand{\rqfourQwenValidCases}{55}
\newcommand{\rqfourQwenPreviousValidCases}{56}
\newcommand{\rqfourTerraValidCases}{56}
\newcommand{\rqfourUnionRecovered}{20}
\newcommand{\rqfourAllRecovered}{7}
\newcommand{\rqfourDeclaredStops}{28}
\newcommand{\rqfourContradictedStops}{3}
\newcommand{\rqfourTerraCost}{14.53}
\newcommand{\solSubsetCases}{15}
\newcommand{\solRecoveredCases}{3}
\newcommand{\solEarlyStopCases}{6}
\newcommand{\solProgressedCases}{6}
\newcommand{\solProbeRecovered}{2}
\newcommand{\solEvaluatedCases}{13}
\newcommand{\solExecutedCases}{12}
\newcommand{\solFinalRecovered}{4}
\newcommand{\solFinalBuilds}{16}

\newcommand{\solCrashReproduced}{3}
\newcommand{\solCalls}{140}
\newcommand{\solTokens}{6.1M}
\newcommand{\solCost}{7.68}

\newcommand{\repairAttemptCount}{6,201}
\newcommand{\applicationFailureCount}{583}
\newcommand{\applicationFailureRate}{9.4}
\newcommand{\applicationFailureCases}{374}
\newcommand{\applicationUnresolvedCases}{253}
\newcommand{\applicationAnchorFailures}{122}
\newcommand{\regexAnchorFailures}{48}
\newcommand{\beforeAnchorFailures}{35}
\newcommand{\afterAnchorFailures}{39}

\newcommand{\repeatedRuleIdentifiers}{136}
\newcommand{\repeatedNoRecoveryProgressions}{575}
\newcommand{\repeatedNoRecoveryIdentifiers}{51}
\newcommand{\goToolchainRecoveredCases}{54}
\newcommand{\rustNightlyRecoveredCases}{22}
\newcommand{\repeatedSuccessfulIdentifiers}{3}

\newcommand{\rqfourQwenOnlyVsTerra}{6}
\newcommand{\rqfourQwenTerraRawP}{0.031}
\newcommand{\rqfourQwenTerraAdjustedP}{0.094}
\newcommand{\rqfourOtherAdjustedP}{0.688}
\newcommand{\rqfourHarnessFailures}{4}
\newcommand{\rqfourQwenTokenStops}{11}
\newcommand{\rqfourQwenPreviousTokenStops}{26}
\newcommand{\rqfourTerraTokenStops}{19}
\newcommand{\rqfourTerraCostPerRecovery}{1.45}

\newcommand{\solSubsetQwenRecovered}{4}
\newcommand{\solSubsetQwenPreviousRecovered}{5}
\newcommand{\solSubsetTerraRecovered}{2}
\newcommand{\solPairwiseMinP}{0.375}

\newcommand{\rqone}{How prevalent is build decay, and how are historical build failures distributed?}
\newcommand{\rqtwo}{Can deterministic repair rules repair decayed builds, and can their derivation be automated?}
\newcommand{\rqthree}{How do reusable rules and per-case agents compare in recovery, cost, and the cases they restore?}
\newcommand{\rqfour}{Where does deterministic repair stop, and what does escalating
      to a per-case agent recover beyond it?}

\newcommand{\pairedCases}{50}
\newcommand{\escalationCases}{60}
\newcommand{\reserveCases}{400}

\newcommand{\pairedEligibleCases}{227}

\pgfplotsset{compat=1.17}
\usetikzlibrary{patterns,arrows.meta,positioning}
\protected\def\sys{\textsc{Relay}}
\newcommand{\code}[1]{\texttt{#1}}
\usepackage[most]{tcolorbox}
\newtcolorbox{rqanswer}[1]{enhanced, breakable=false,
  colback=black!5, colframe=black!70, boxrule=0.8pt, arc=2.5pt,
  left=6pt, right=6pt, top=6pt, bottom=5pt,
  before skip=4pt, after skip=6pt,
  drop shadow={black!50, opacity=0.35},
  fontupper=\raggedright,
  title={Answer to #1}, fonttitle=\normalsize, coltitle=white,
  attach boxed title to top left={xshift=4pt, yshift=-7pt},
  boxed title style={colback=black!55, colframe=black!70, boxrule=0.8pt,
    arc=2.5pt, left=4pt, right=4pt, top=1.5pt, bottom=1.5pt}}
\begin{document}
\title{RELAY: Scaling Historical Build Restoration through Rule Reuse and Agentic Repair}
\author{Olivier Nourry}
\affiliation{%
  \institution{The University of Osaka}
  \city{Osaka}
  \country{Japan}}
\email{nourry@ist.osaka-u.ac.jp}

\author{Yoshiki Higo}
\affiliation{%
  \institution{The University of Osaka}
  \city{Osaka}
  \country{Japan}}
\email{higo@ist.osaka-u.ac.jp}

\begin{abstract}
Vulnerability benchmarks and software engineering research depend on rebuilding historical software. Yet builds rely on external resources, such as base images and packages, that change independently of the source code. In this study, we examine this environment decay on \totalBuilds{} OSS-Fuzz cases. We find that \failRate\% no longer complete reconstruction and verification, and that most failures share a small number of recurring failure symptoms. A recurring failure can therefore be diagnosed once and its repair reused. We present \sys{}, a hybrid framework that uses an LLM to derive and validate deterministic repair rules from representative failures, reuses them without model calls, and escalates cases unresolved by rules to a per-case agent. We evaluate \sys{} in human-guided and fully automated rule-derivation campaigns and compare it with a per-case agent approach. We find that rule derivation can be fully automated, and that reusable rules restore as many builds as per-case agents with fewer build executions. However, agents restore cases that the rules cannot. Our results show that a hybrid approach that applies rules first and escalates the remaining cases to an agent restores more builds than either approach alone and at a lower cost than using agents alone.
\end{abstract}



\maketitle

\section{Introduction}
\label{sec:intro}
Rebuilding an earlier version of a software system is necessary for
reproducing reported bugs, bisecting regressions, and evaluating tools on
historical defects. This need is especially visible in vulnerability research.
For instance, ARVO reconstructs vulnerable and fixed versions of OSS-Fuzz
projects~\cite{arvo} and benchmarks such as CyberGym and SEC-bench depend on
executable historical software~\cite{cybergym,secbench}. If a case no longer
builds, its vulnerability and any proposed repair cannot be evaluated through
execution. Researchers must instead exclude the case or rely on indirect
measures such as code similarity, which may not reflect whether a patch
actually fixes the vulnerability~\cite{coderovers}.

Rebuilding historical software is often much harder than checking out its
source code. Build instructions may remain available, but they depend on
resources that change independently of the target project. Base images are
retired, package repositories remove old versions, toolchain installers move,
and third-party download URLs disappear~\cite{nourry2024,arvo,malka2026docker}. A build can
therefore fail years later even though its target source revision has not
changed. We refer to failures caused by such changes as \emph{environment
decay}. Developers generally maintain the current version of a project rather
than restore every historical release, leaving this work to the researchers and
benchmark maintainers who depend on those releases.

Existing approaches repair build environments through reusable operations,
such as Shipwright's Dockerfile repairs~\cite{henkel2021shipwright}, or through
per-case language-model agents~\cite{repo2run2025,executionagent2025}. Per-case agents are flexible, but they repeat the diagnosis, patch generation,
and validation process for every build. This is inefficient when the same
environment failure affects many builds, since each case requires new model
calls and one or more build executions. Moreover, external dependencies
continue to change, so a restored build may fail again later. A scalable
approach should therefore reuse previously validated repairs where possible,
while retaining per-case reasoning for failures that cannot be handled by
existing rules.

In this study, we use historical OSS-Fuzz cases to examine whether reusable
rules and per-case agent escalation can restore build environments efficiently
at scale. These cases cover different projects, languages, dependency managers,
and toolchains. Their recorded crashing inputs also let us evaluate whether a
repaired environment preserves the expected historical behavior, rather than
measuring build completion alone.

We investigate a hybrid approach to historical build restoration. We implement
\sys{}, a framework built around ARVO's historical reconstruction pipeline.
\sys{} converts build logs into normalized failure observations, groups related
failures, and selects representative cases. An LLM uses evidence from these
cases to propose a repair rule. The framework checks the permitted repair
scope, applies the rule, and validates it by rebuilding the affected cases.
Validated rules are stored in a registry and reused whenever their conditions
hold. When a repair exposes a different failure, \sys{} records the new
evidence and continues from the updated build state. Cases that remain
unresolved are escalated to a per-case agent.

We aim to produce rules that recognize recurring failures and express repairs that can
apply across cases. A rule may pin a stable version when that version addresses
a recurring compatibility or availability problem, but it should not simply
encode a one-off change for a single project. The model proposes both the
conditions under which a rule applies and the changes it makes. For example,
several builds may fail because they refer to a retired source archive. A rule
should recognize the relevant download failure and redirect the request to a
supported mirror or archive without reasoning about every affected case
separately.

Our study separates the prevalence of environment decay from the evaluation of
repair. We first examine \totalBuilds{} historical OSS-Fuzz cases with recorded
execution outcomes and find that \usableBuilds{} do not complete reconstruction
and verification. We characterize the failures using the \usedBuilds{} cases
for which we collected detailed failure logs. We then evaluate whether rules derived from representative failures transfer to
other cases and whether their derivation can be automated. We compare rules,
per-case agents, and their rule-first combination on the same failures.
Finally, we test agent escalation on a separate set of cases where deterministic
repair had stopped.

The paper addresses four questions:
\begin{enumerate}
\item \textbf{RQ1: \rqone}
Reconstruction failed for \failRate{}\% of the \totalBuilds{} cases we
executed. The failures were concentrated: \concGroups{} groups contained
at least five cases and represented \concShare{}\% of the failure-analysis dataset. We
found no consistent relationship between build age and repair outcome, but
outcomes varied substantially across projects.
\item \textbf{RQ2: \rqtwo}
LLM-derived rules contributed to \ruleRecoveries{} restored builds. The
automated campaign contributed \autoRecoveries{} recoveries in two weeks,
compared with \humanRecoveries{} during three months of human-guided work.
\item \textbf{RQ3: \rqthree}
Rules and per-case agents restored similar numbers of builds, but the
agents used \agentBuildLo{} to \agentBuildHi{} times as many build executions
per recovery. They also restored different cases. Combining their recorded
outcomes restored \hybridTerraRecovered{} to \hybridQwenRecovered{} of the
\pairedCases{} cases.
\item \textbf{RQ4: \rqfour}
Rules stopped when their edits did not fit a project's build files or when
derivation repeated variations of an unsuccessful strategy. On a subset of
\escalationEvaluatedCases{} cases that remain unresolved after rule-based repair, agents restored
\rqfourTerraRate{} to \rqfourQwenRate{}\%.
\end{enumerate}

This paper makes the following contributions:
\begin{itemize}
  \item We conduct a large-scale investigation of environment decay in \totalBuilds{} historical OSS-Fuzz cases and show that build decay is widespread and that failure causes tend to recur across unrelated cases.
  \item We present \sys{}, a scalable framework that derives reusable deterministic
  repair rules from representative failures, validates them through execution,
  and escalates unresolved cases to a per-case agent. It also extends historical
  build verification beyond C and C++.
  \item We show that the framework can derive and validate reusable repair rules without human intervention and use them to restore historical builds.
  \item We compare rule-based and per-case agentic build repair and show that a hybrid strategy combining the cost-efficiency of reusable rules with the flexibility of per-case agentic repairs is effective for large-scale build restoration.
  \item We provide a replication package with the rule registry, failure observations, repair records, prompts and model responses, and analysis scripts. We provide a partial version for review and will release the full package upon publication.
\end{itemize}

\section{Background}
\label{sec:background}
\subsection{OSS-Fuzz and Historical Builds}
OSS-Fuzz is a continuous fuzzing platform that fuzzes open-source software to detect crashes and other
incorrect behavior~\cite{serebryany2017ossfuzz}. When it confirms a failure, it creates a report identifying the affected project,
fuzz target, sanitizer configuration, and input that triggered the
failure~\cite{arvo}. A reported failure shows that the project built and that its fuzz target executed successfully when the issue was discovered.

Projects integrated with OSS-Fuzz share a common configuration structure. Each
project supplies a \code{project.yaml} file containing project metadata, a
\code{Dockerfile} defining its build environment, and a \code{build.sh} script
that compiles the project and its fuzz targets inside that
environment~\cite{ossfuzzguide}. This common structure provides a consistent
interface for executing builds, but it does not make those builds
self-contained. A Dockerfile may download base images, install system packages,
or retrieve source archives, while the build script may resolve language
packages, invoke code generators, and select a compiler or toolchain. These
external resources can change or disappear even when the project files and
target source revision remain unchanged.

\subsection{Executable Vulnerability Datasets and ARVO}

Executable historical software is increasingly used to evaluate vulnerability
discovery and repair systems. CyberGym evaluates vulnerability
reproduction, while SEC-bench evaluates both proof-of-concept generation and
vulnerability patching~\cite{cybergym,secbench}. These benchmarks
require more than a source snapshot: the project must build, the
relevant target must execute, and the recorded input must still exercise the
expected behavior.

ARVO reconstructs vulnerable and fixed versions of vulnerabilities reported by
OSS-Fuzz~\cite{arvo}. It combines information from the OSS-Fuzz report with
historical project and dependency revisions, the OSS-Fuzz integration files,
and the recorded crashing input. ARVO also addresses several recurring
obstacles to historical reconstruction, including dependency revision
selection, base-environment changes, and unavailable external resources. ARVO's published dataset and evaluation focus on C and C++ memory-safety
vulnerabilities. While its reconstruction approach can be extended to other
ecosystems, doing so requires language-specific handling of dependencies,
build outputs, and crash behavior. We extend the workflow to Rust, Go, Python,
and JVM projects and add the necessary logging and validation needed to distinguish build
completion from successful reproduction of the historical vulnerability. The
repairs studied in this paper begin with the failures that remain after ARVO's
existing reconstruction mechanisms have run.

\subsection{Build and Environment Repair}
Build failures can arise from source errors, dependency changes, unavailable
resources, incompatible tools, or faults in build and container
configuration. Existing repair systems often address one part of this space.
For example, Shipwright groups Dockerfile failures and develops reusable
repairs through a human-in-the-loop process~\cite{henkel2021shipwright}.
Other approaches repair dependency-related build failures or restore
language-specific environments~\cite{macho2018automatically,
mukherjee2021pydfix}. More recent work includes language-model agents that
configure and execute repositories individually~\cite{repo2run2025,
executionagent2025}, as well as EnvBench, a benchmark for automated environment
setup~\cite{envbench2025}.

Prior work on OSS-Fuzz build failures found 25 causes of fuzzing build
failures across project source, build scripts, dependencies, toolchains,
external resources, and execution infrastructure~\cite{nourry2024}. These
failures can occur in sequence; repairing a Docker image may expose a dependency
error, and repairing that dependency may expose a compiler incompatibility.
Success in progressing past one build error therefore does not establish that the complete historical build has been restored.

\sys{} combines the cost-efficiency of reusable rules with the flexibility of per-case agentic repair. It uses an LLM to
derive deterministic rules from representative failures and applies validated
rules to matching cases without further model calls. Cases that remain
unresolved are escalated to a per-case agent. Both repair paths (rule and agentic) may modify
files at different stages of the build, but they must preserve the historical
target revision and recorded crashing input.

\section{\sys{} Framework}
\label{sec:approach}
\begin{figure*}[t]
\centering
\includegraphics[width=\textwidth]{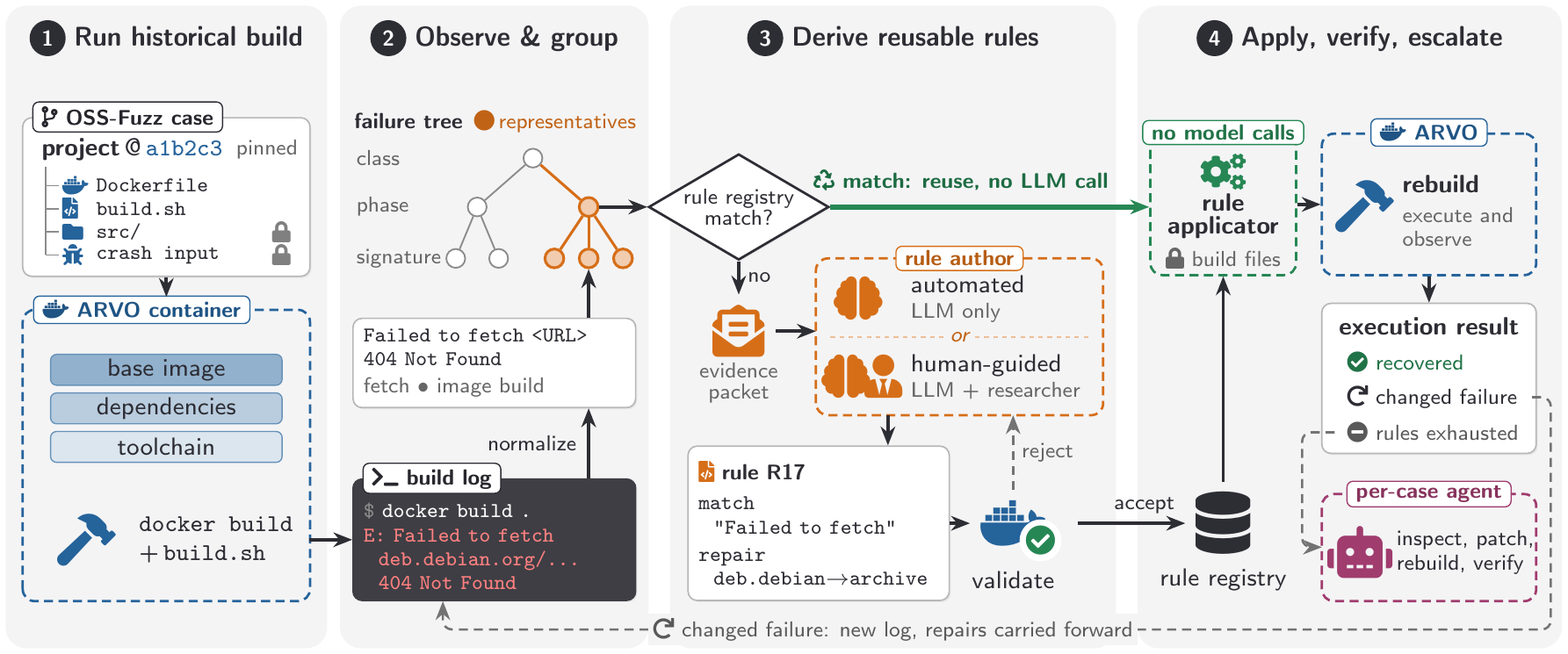}
\caption{The \sys{} repair pipeline. (1)~Each historical OSS-Fuzz case is
built in an ARVO container; target source and the crashing input are
protected. (2)~The build log is normalized into a signature and placed in a
failure tree whose high-yield branches supply representative cases. (3)~If an existing
rule matches, it is reused without a model call; otherwise a bounded evidence
packet enters the rule-authoring process, where a model is invoked automatically
or under researcher guidance. The proposed rule enters the registry only after
it rebuilds the representative cases. (4)~Rules are applied deterministically to build-environment files. A changed
failure is rebuilt with the repair chain carried forward. Eligible cases that
remain unresolved are escalated to a separately budgeted per-case agent, while
cases blocked by unavailable resources or execution infrastructure are
deferred.}
\label{fig:pipeline}
\end{figure*}

Figure~\ref{fig:pipeline} shows how \sys{} combines deterministic repair, LLM-assisted rule derivation, and per-case agent escalation. The framework
reconstructs builds, normalizes and groups failures, selects representative
cases, prepares evidence, matches and applies rules, executes builds, and
records outcomes. A rule-authoring LLM operates on groups of related failures
and proposes deterministic rules intended for reuse. A per-case repair agent is
invoked only after deterministic repair stops. It may inspect and modify one
case iteratively, but remains subject to the same protected-file policy and to
explicit limits on turns, builds, and tokens.

\subsection{ARVO Integration and Repair Scope}
\label{sec:integration}

\sys{} uses ARVO as its execution mechanism for reconstructing a project at a
specific historical revision. ARVO already controls parts of the base
environment, dependency revisions, and resource retrieval, and attempts its own
environment fixes during reconstruction~\cite{arvo}. Our repair loop therefore begins with
the failures that remain after those mechanisms have run. It surrounds ARVO
with evidence collection, reusable rule derivation and application, repair
state management, and agent escalation. \sys{} does not replace ARVO's
reconstruction workflow but rather invokes that workflow repeatedly to observe and validate environment repairs.

The original ARVO workflow used as our starting point selects C/C++ cases. We
extend selection and verification to non-C/C++ projects, including language-aware
handling of Rust, Python, and JVM crash signals and target wrappers. The
Dockerfile/build-script interface made much of the existing build machinery reusable
across ecosystems, while crash verification required separate handling.
We also add an interface for applying repair rules and record which repairs were attempted, whether they were applied, and how each execution ended. These records let us distinguish project build failures, repairs that could not be applied, and failures in the execution infrastructure.

To reduce the risk of modifying program behavior, we permit repairs only to the
build environment: Dockerfiles, build scripts, package-manager configuration,
dependency metadata, resource locations, and toolchain setup. The target source
revision, program source, crashing input, fuzz target, and sanitizer
configuration are protected. An attempted change to a protected input, whether
produced by a rule or a repair agent, is rejected before execution.

\subsection{Observations, Grouping, and Evidence}
Each execution produces a run directory containing output streams, status
records, available log metadata, and a normalized case observation. \sys{} classifies the failure, estimates the execution phase reached at which it occurred, and derives a signature from the relevant log excerpt. \sys{} selects excerpts using generic and ecosystem-specific failure markers and the evidence requirements of existing rules. It then normalizes dynamic
content such as paths, identifiers, hashes, version numbers, line numbers, and
repeated whitespace. The grouping key combines failure class, phase, and
signature hash. Including the phase prevents a similar message during image
creation and target execution from being treated as the same blocker.

\sys{} groups similar failures to select cases for rule derivation, but the same error message can have different causes. For example, removing version numbers during normalization can group failures involving different dependency versions. We therefore keep the original logs for rule derivation and test each rule by rebuilding every case to which it is applied.

\sys{} then organizes observations in a tree from coarse categories to exact signatures:
failure class, phase, cluster family, ecosystem, error pattern, abstract
template, and normalized signature. This hierarchy lets the framework
prioritize high-yield branches, where one reusable repair may affect many
builds, as well as branches that have not yet been examined. We refer to a case
selected from a branch for rule authoring and initial validation as a
\emph{representative}. Representatives provide concrete evidence for the
branch, but do not eliminate the need to validate other matching cases.

Rather than provide complete logs to the rule-authoring model, \sys{}
constructs bounded evidence packets. A packet contains relevant excerpts, the
failure class, phase and signature, case metadata, and repair history. The
history records previously applied rules and how the observed failure changed
after each rebuild. Excerpt selection uses generic and ecosystem-specific
failure markers, the evidence requirements of existing rules, and removes repeated or redundant log content. \sys{} records packet sizes and token counts to bound the
cost of each proposal, with the trade-off that a packet may omit useful
context. When the available evidence is insufficient, the framework schedules
a fresh execution or a targeted diagnostic execution and constructs a new packet before requesting
another proposal.

\subsection{LLM-Assisted Rule Derivation}

\begin{figure}[!htb]
\includegraphics[width=0.7\linewidth]{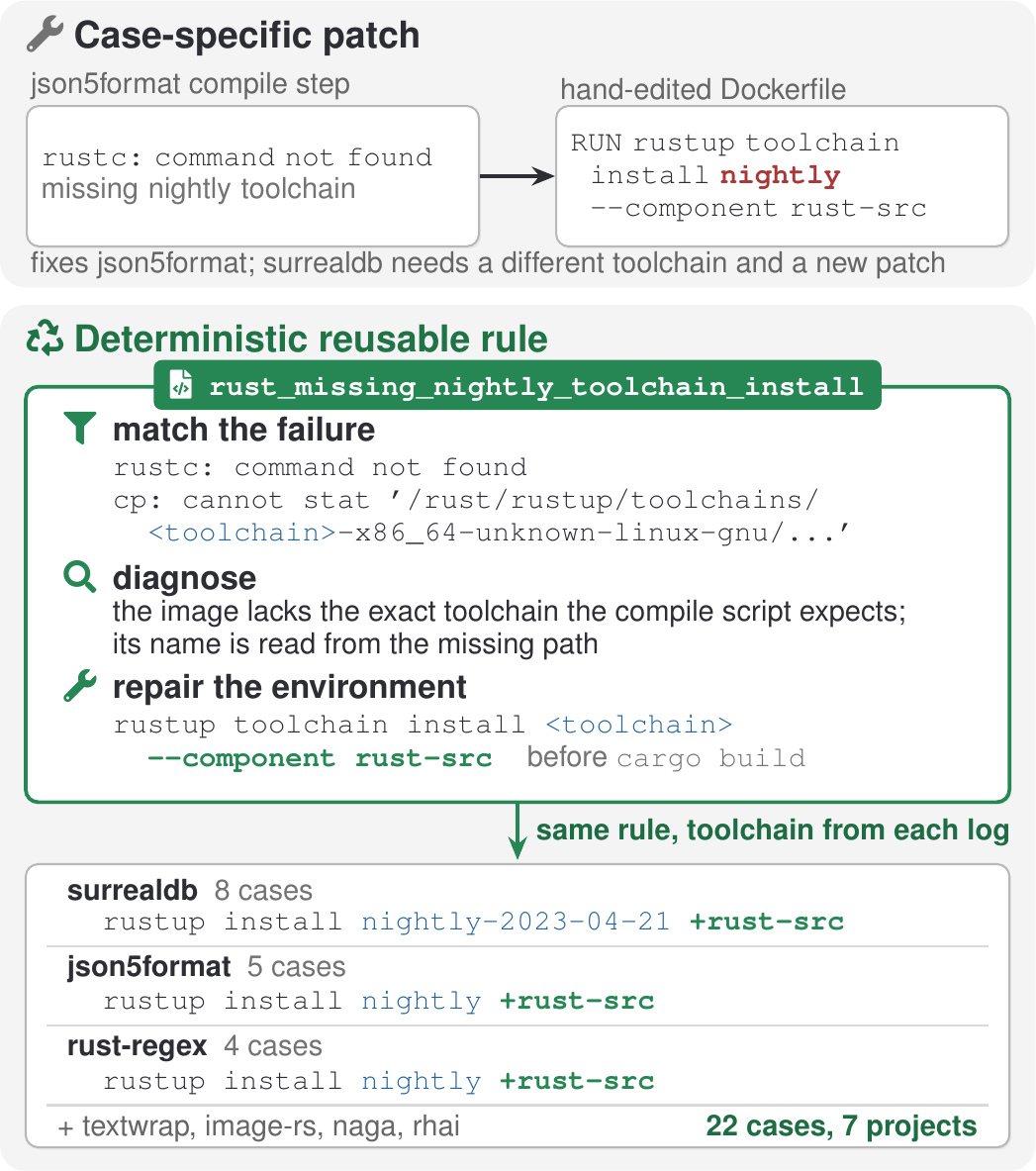}
\caption{Difference between a case-specific patch and a reusable rule. The
patch hard-codes one toolchain for one project. The rule reads the missing
toolchain from the failing path and installs it with its source component,
recovering 22 cases across 7 projects with different toolchains.}
\label{fig:rule_definition}
\end{figure}

Before requesting a new rule, \sys{} checks the current observation against the
validated rules in its registry. If a rule matches, the framework applies it
without invoking a model. Rule derivation begins only when the registry contains
no eligible rule for the observed failure.

For a selected branch, the rule-authoring LLM receives evidence packets from
its representatives, relevant repair history, and descriptions of existing
rules. It returns a structured candidate containing a failure precondition,
parameters to extract from the evidence, permitted repair operations, evidence
requirements, and allowed file scope. We instruct the model to express repairs
in terms of the observed failure rather than project names, case identifiers,
or constants copied from one case. For example, Figure~\ref{fig:rule_definition}
shows a toolchain rule that reads the name of the missing Rust toolchain from
the failing path and installs it, instead of hard-coding one toolchain. A version may be pinned when the evidence establishes a
compatibility requirement, but the rule must also state when that pin applies.

In the human-guided campaign, one of the authors selected branches, reviewed the
evidence, and guided or revised model proposals. In the fully automated
campaign, the framework selected branches, constructed prompts, invoked the
model, materialized its proposal, and scheduled validation without human
intervention. In both settings, the model proposed a repair, but execution
determined whether it worked.

A proposal is not immediately added to the rule registry. The framework checks
its scope, applies its operations against the historical build files, and
rebuilds representative cases. Before adding a proposal to the registry, the framework checks its scope, applies its operations to the historical build files, and rebuilds representative cases. It accepts the rule only if the execution restores a build or exposes a different failure. Once a rule is adopted, later applications require no further model calls.

\subsection{Rule Matching and Deterministic Application}

A rule pairs a failure precondition with a repair specification. Preconditions
describe the observation and relevant context. Repair specifications use
bounded operations such as rewriting a URL, pinning a dependency, installing a
package, selecting a toolchain, replacing a base image, or editing a build
script. The \emph{rule applicator} applies these operations against the
historical build files.

Three decisions remain separate. First, the matcher determines whether the
current observation satisfies a rule's preconditions. Second, the applicator
determines whether the requested operations can be carried out on the available
files without violating the repair scope. Third, a new ARVO execution determines
whether the applied repair recovers the build, exposes a different failure, or
leaves the original failure unchanged. We record all three outcomes so that a
matching error, an application failure, and an ineffective repair are not
treated as the same event.

Each rule targets a specific \emph{breakage step}: an identifiable operation that
fails during reconstruction, such as retrieving a retired archive, installing
a missing package, or invoking an incompatible toolchain. The same rule can repair a failure across several versions or projects if it recognizes the failure and can apply the required change to their build files.

A build may fail again after its first error is fixed. For example, an unavailable base image prevents compilation from starting, while a failed source download can hide a later dependency error. \sys{} therefore keeps edits that resolve an intermediate failure and rebuilds after each applied rule. These edits form a repair chain. After each rebuild, \sys{} uses the latest failure to select the next rule. Cases that initially failed for different reasons may then encounter the same failure and reuse the same rule.

\sys{} determines whether a repair restored the build or exposed a different failure by rebuilding the case. It records a change in failure class or signature as a \emph{progression}, rather than a recovery. When this happens, the framework keeps the repair chain and continues from the newly observed failure.

\subsection{Stopping and Agent Escalation}

\sys{} stops deterministic repair when the build recovers, no eligible rule remains, two consecutive attempts return the same failure state, or the required action exceeds the permitted scope. It records why it stopped, including whether a case needs a new rule, an unavailable resource, case-specific reasoning, or an edit that the applicator could not apply. It records infrastructure failures and provenance problems separately from repair outcomes.

\sys{} escalates cases with a repairable build failure when no applicable rule remains. It defers cases blocked by infrastructure failures or a confirmed unavailable resource. The agent receives the case’s current failure, complete repair history, and build files. It can inspect logs and files, probe the environment, make edits, and invoke ARVO. It keeps earlier edits as it works on subsequent failures.

The agent aims to repair one case. Its patch does not need to generalize, and \sys{} does not automatically add it to the rule registry. The agent can make case-specific edits that the deterministic rule operations cannot express, but it cannot change the protected source revision, program source, fuzz target, crashing input, or sanitizer configuration. \sys{} limits the model turns, build executions, and tokens available for each case and counts a recovery only when a new execution confirms that the build has been restored.

\section{Study Setup}
\label{sec:design}
We answer four research questions using different subsets of the OSS-Fuzz cases
described below. RQ1 measures the prevalence and distribution of historical
reconstruction failures. RQ2 evaluates reusable repair rules and whether their
derivation can be automated. RQ3 compares reusable rules, per-case agents, and
their rule-first combination on the same failures. RQ4 examines why
deterministic repair stops and what per-case agents recover after it does.

\subsection{Dataset and Study Subsets}
\label{subsec:dataset}

We constructed the dataset from an OSS-Fuzz issue-tracker snapshot collected on \trackerSnapshotDate{}. Each eligible issue contributes at most one case containing the project, fuzz target, crashing input, sanitizer, and historical source revisions required by ARVO. Issues without sufficient reconstruction metadata are excluded.

Historical reconstruction is computationally expensive: each execution attempts
to restore the build environment, build the relevant project revisions, and run
the recorded crashing input. Across the cases in our dataset, the recorded
execution times total approximately \recordedExecutionHours{} hours, and
\hourLongExecutions{} cases took at least one hour. We therefore evaluated a
broad subset of the collected issues rather than attempting every issue in the
tracker.

C and C++ account for \cppIssueShare\% of the \trackerIssues{} issues in our
metadata. To include failures from other ecosystems, we executed the available
cases from projects identified as using other languages and then sampled C/C++
cases in random order. After removing repeated case identifiers, the dataset
contains \totalBuilds{} distinct historical cases. Table~\ref{tab:population}
summarizes the dataset and the subsets used in each evaluation.

Our pipeline first invokes ARVO’s existing reconstruction mechanisms, including its handling of historical dependencies, base environments, and unavailable resources. The failures studied here are therefore cases that remain unresolved after this initial reconstruction. Of the \totalBuilds{} cases in the dataset, \usableBuilds{} did not complete reconstruction and verification, giving a failure rate of \failRate\%. ARVO’s published reproduction rate~\cite{arvo} and the failure rate reported in this paper come from different case selections and execution periods and should not be compared directly. 

\begin{table}[t]
\caption{Dataset and evaluation subsets used in the study.}
\label{tab:population}
\centering
\small
\begin{tabular}{@{}lr@{}}
\toprule
Dataset or evaluation subset & Cases \\
\midrule
Historical cases executed & \totalBuilds{} \\
Cases that failed reconstruction & \usableBuilds{} \\
Failed cases with usable failure evidence & \usedBuilds{} \\
Cases reached by the repair campaigns & \attemptedBuilds{} \\
\midrule
Cases held out from rule derivation (RQ3) & \reserveCases{} \\
Cases still failing before the rule-agent comparison (RQ3) & \pairedCases{} \\
Cases selected for agent evaluation (RQ4) & \escalationCases{} \\
Cases still failing when re-executed (RQ4) & \escalationEvaluatedCases{} \\
\bottomrule
\end{tabular}
\end{table}

The first four rows show how we narrowed the dataset from all executed cases to those attempted during the repair campaigns. We use all \totalBuilds{} executed cases to measure the prevalence of build failures. We then collected detailed execution logs for \usedBuilds{} cases and assembled them into a failure-analysis dataset to classify and group failures. Out of these, we attempted to repair \attemptedBuilds{} cases during the repair campaigns. We use
three separately selected groups for the controlled evaluations. We held out \reserveCases{} previously unattempted cases to test whether our rule registry could repair failures beyond the cases explored during the campaign. We also selected \pairedCases{} previously repaired
cases that still fail on re-execution to compare reusable rules with per-case
agents on the same inputs. Lastly, we selected \escalationCases{} cases
recorded as unresolved after deterministic repair. Before running the agents,
we rebuilt each case without a repair to verify that they were still failing and found that \escalationProbeRecovered{} cases built successfully. Their outcomes had therefore changed between the time we sampled them and
the time we evaluated them which can occur for various reasons such as external dependencies not being fully pinned.
We excluded them because they no longer presented a build failure and used the
remaining \escalationEvaluatedCases{} cases to test what per-case agents
recover after rule-based repair stops.

The failure-analysis dataset contains \cppLogCases{} C++ cases,
\rustLogCases{} Rust, \goLogCases{} Go, \jvmLogCases{} JVM, \cLogCases{} C,
\pythonLogCases{} Python, and \swiftLogCases{} Swift. Non-C/C++ ecosystems
account for \nonCppLogCases{} cases, or \nonCppLogPct\% of the dataset. The
large C/C++ share is expected because C and C++ projects make up most of
OSS-Fuzz~\cite{shirai2026language}. We retain Swift in the overall totals but
do not report a separate ecosystem-level rate for its small group.

\subsection{Repair Selection and Budgets}
\sys{} does not try to repair every case separately. It groups similar failures and selects representatives from promising groups. It then derives and validates rules on those representatives. A case is executed when it is chosen as a representative or when an accepted rule matches its failure. This gives priority to recurring failures for which one rule may repair several cases.

Across the RQ1 and RQ2 campaigns, \sys{} attempted to repair \attemptedBuilds{} distinct cases. It rebuilt each case after applying a
repair, sometimes exposing another failure that required a further change. \sys{} stopped deterministic repair when the build recovered or when two consecutive repair
attempts returned the same failure state without exposing a new one. We set limits on build executions, model turns, tokens, and cost for agent runs, as summarized in
Table~\ref{tab:settings}. Note that an unresolved case may still be repairable with a larger budget.

\subsection{Per-Case Agent Configuration}
\label{subsec:agentconfig}
All models use the same agent harness. Each agent receives the case evidence,
complete build logs, and historical build files. It can inspect files, run
read-only shell commands, edit the permitted build-environment files, and
rebuild the case. The harness disables web access, and prevents changes to the protected source revision,
program source, fuzz target, crashing input, and sanitizer configuration. After each rebuild, the agent
receives the latest failure and can address the newly exposed blocker without
discarding earlier changes. If the harness cannot apply a patch, it returns
the error to the agent so that it can propose a correction.

Table~\ref{tab:settings} summarizes the agent configuration. All models use the
same limits on build executions, turns, and total tokens per case. The two Qwen
models also use the same context size and sampling settings. The harness keeps each model's reasoning and conversation history between turns. The token limit includes all input and output across model calls. In both evaluations, we ran each model once on each assigned case.

\begin{table}[t]
\caption{Agent settings. Local sampling and context settings apply to both
Qwen releases; hosted sampling uses the API default.}
\label{tab:settings}
\centering\footnotesize
\begin{tabular}{@{}p{0.36\columnwidth}p{0.59\columnwidth}@{}}
\toprule
Setting & Value \\
\midrule
Local models & Qwen3.8-27B and Qwen3.6-27B, Q4\_K\_M \\
{Local serving} & Ollama, four one-GPU replicas (RTX A6000, 48 GB) \\
Local sampling & Temperature 1.0, top-$p$ 0.95, top-$k$ 20 \\
Local context & 131,072 tokens \\
Hosted models & Terra: medium; Sol: high reasoning effort \\
Reasoning & Enabled and preserved across turns \\
Output limit per turn & 16,384 tokens \\
Per-case limits & 5 builds; 70 turns; 600,000 total tokens \\
Unchanged-failure stop & Two consecutive builds \\
Invalid-response stop & Three consecutive invalid responses \\
Hosted cost ceiling & Terra: \$2/case; Sol: \$5/case \\
Execution & One build per case at a time; at most four cases in parallel \\
\bottomrule
\end{tabular}
\end{table}

\subsection{Outcome and Cost Measures}
\label{subsec:outcomes}
Our primary outcome is build recovery (i.e., the build completes
successfully). We also check whether the repaired environment preserves the historical crash behavior. We count the vulnerability as reproduced when the recorded input triggers a crash at the vulnerable revision. We classify a case as \emph{benchmark-ready} when the input triggers the crash at the vulnerable
revision and runs without a crash at the designated fixed revision.

We measure repair effort through build executions. For each approach, we
divide the total number of build executions across all evaluated cases by
the number of recovered cases. This includes executions spent on cases
that remain unresolved. Proposed edits that the harness cannot apply do
not launch a build and do not contribute to this count. For agent runs,
we also report model calls, token use, and hosted-model charges.

We inspect execution errors before classifying outcomes. We count patches
that fail to apply, attempts to edit protected files, and decisions by the
model to stop as agent outcomes. We report results for all selected cases.
For each paired comparison, we also report results after excluding cases
where either approach encountered a harness failure that prevented a
reliable outcome.

\section{RQ1: \rqone}
\label{sec:rq1}
\subsection{Motivation}
Before developing a repair framework, we first need to establish whether historical build failures are common and whether the same failures affect multiple cases. Reusable rules are useful only when failures repeat. RQ1 therefore measures how often historical reconstruction fails, how strongly the failures are concentrated, and whether repair outcomes vary by age or project.

\subsection{Approach}
We calculated the overall failure rate across the \totalBuilds{} cases we executed. We used the \usedBuilds{} cases in the failure-analysis dataset to examine how often the same failure occurred across multiple cases.

We assigned each case to the first failure observed before any repair was
attempted. As described in Section~\ref{sec:approach}, we represented each
failure by its class, execution phase, and normalized signature. Cases sharing
all three values formed a baseline failure group. We report the number and
size of these groups, along with the share of cases in groups of at least five
and the share that occurred alone. We chose five as the threshold because it
left enough cases to inspect representatives and test whether a repair
transferred to other cases. We kept this initial grouping fixed throughout the
analysis. Failures exposed by later repair attempts did not change a case's
baseline group.

We also measured how often a failed case later built without a repair. Among
the \attemptedBuilds{} cases included in the repair campaigns, we classified a
case as a \emph{no-patch recovery} when a later execution succeeded without a repair
chain. This showed that re-execution was sufficient, but not why the initial
execution failed. For the age and project analyses, we used outcomes from the
deterministic repair campaigns evaluated in RQ2. We classified a case as \emph{recovered}
if any campaign execution succeeded, with or without a repair. Of the \ageDatedCases{} attempted cases with a known vulnerable revision
date, we included those from projects with at least ten such cases
spanning at least two historical years.
These criteria left \ageEligibleCases{} cases from \ageComparedProjects{}
projects and gave each half of the within-project comparison at least five
cases. Within each project, we ordered cases by their vulnerable revision date
and split them at the median. We then compared the unresolved rates of the
older and newer halves. We used an exact two-sided sign test, excluding ties,
to determine whether the differences consistently pointed in either direction.
Finally, we report project-level unresolved rates only for projects with at least
twenty attempted cases to avoid comparisons based on only a few cases.

\subsection{Results}
\textbf{Historical reconstruction frequently fails.} Of the
\totalBuilds{} cases we executed, \usableBuilds{} (\failRate{}\%) stopped before
ARVO completed reconstruction and verification. Although their source revisions
were pinned, these cases failed for various reasons, including project build failures, unavailable resources, and problems in the execution infrastructure.

\begin{figure}[t]
\centering
\begin{tikzpicture}
\begin{axis}[
  width=0.72\linewidth, height=2.3in,
  xmode=log, log basis x=10, xmin=1, xmax=1106,
  ymin=0, ymax=100,
  xtick={1,10,100,1000}, xticklabels={1,10,100,1{,}000},
  ytick={0,20,40,60,80,100},
  xlabel={Largest failure groups included (log scale)},
  ylabel={Cases covered (\%)},
  axis x line*=bottom, axis y line*=left,
  tick label style={font=\footnotesize},
  label style={font=\footnotesize},
  clip=false,
]
\addplot[black, line width=1pt, mark=none] coordinates {
(1,7.62) (2,13.36) (3,16.69) (4,19.74) (5,22.49) (6,24.8) (7,26.79) (8,28.23)
(9,29.54) (10,30.62) (11,31.66) (12,32.62) (13,33.55) (14,34.37) (15,35.19)
(16,35.96) (17,36.67) (18,37.36) (19,38.05) (20,38.71) (21,39.38) (23,40.66)
(24,41.26) (25,41.83) (26,42.37) (29,43.9) (30,44.38) (31,44.85) (33,45.73)
(39,48.26) (41,49.06) (43,49.81) (44,50.17) (45,50.5) (47,51.12) (54,53.14)
(60,54.73) (70,57.17) (79,59.16) (87,60.76) (100,63.06) (118,65.85)
(135,68.11) (161,70.99) (208,75.16) (260,78.62) (379,83.89) (1106,100.0)};
\draw[dashed, black!40] (axis cs:10,0) -- (axis cs:10,30.62) -- (axis cs:1,30.62);
\draw[dashed, black!40] (axis cs:50,0) -- (axis cs:50,52.0) -- (axis cs:1,52.0);
\draw[dashed, black!40] (axis cs:161,0) -- (axis cs:161,70.99) -- (axis cs:1,70.99);
\fill (axis cs:10,30.62) circle (1.7pt);
\fill (axis cs:50,52.0) circle (1.7pt);
\fill (axis cs:161,70.99) circle (1.7pt);
\node[anchor=south east, font=\scriptsize, inner sep=2pt] at (axis cs:10,30.62) {10 groups: 30.6\%};
\node[anchor=south east, font=\scriptsize, inner sep=2pt] at (axis cs:50,52.0) {50 groups: 52.0\%};
\node[anchor=south east, font=\scriptsize, inner sep=2pt, align=right] at (axis cs:161,70.99) {161 groups with at least\\five cases: 71.0\%};
\draw[black!60, |-|] (axis cs:380,74) -- (axis cs:1106,74)
  node[midway, below, font=\scriptsize, align=center, text=black!75] {727 single-case\\groups: 16.1\%};
\end{axis}
\end{tikzpicture}
\caption{Cumulative share of the \usedBuilds{} cases in the failure-analysis
dataset covered by the largest failure groups. The ten largest groups cover
\topTenCasePct{}\% of the cases, and the \concGroups{} groups with at least five
cases cover \concShare{}\% of all cases.}
\label{fig:concentration}
\end{figure}
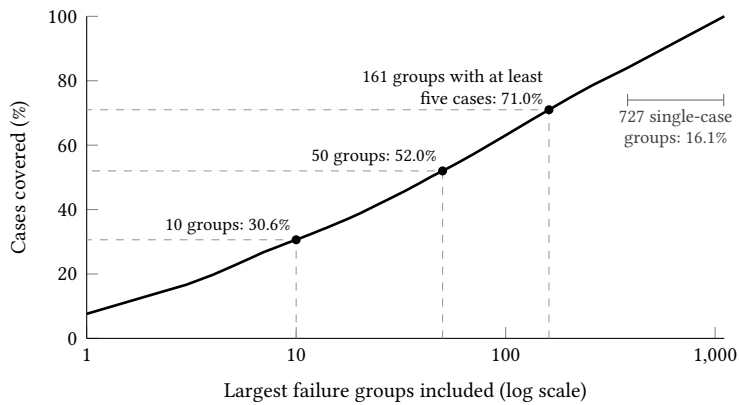

\textbf{Initial failures are highly concentrated.} The \usedBuilds{} cases in
the failure-analysis dataset formed \baselineGroups{} failure groups.
Figure~\ref{fig:concentration} ranks these groups by size. The ten largest
groups covered \topTenCasePct{}\% of the cases. The \concGroups{} groups with at
least five cases covered \concShare{}\%, whereas the \singletonGroups{}
single-case groups account for \singletonCasePct{}\%. Many cases therefore shared
the same initial diagnosis.

These groups identify promising opportunities for reusable repair rules, though a repair is not guaranteed to work for every case in a group.
Cases with the same initial failure may expose different blockers after that
failure is removed. Conversely, a case that initially occurs alone may later
match a rule derived from another group. We therefore use failure groups to
identify candidates for reuse and test whether the same repair works on other cases.

\textbf{Some failures disappear on re-execution.} Among the
\attemptedBuilds{} attempted cases, \noPatchRecoveries{} (\noPatchRate{}\%)
later built without a repair. We could not determine
why these cases succeeded on a retry. These recoveries prompted us to test
whether builds that succeeded after applying rules actually needed those
repairs. We report this evaluation in RQ2.

\textbf{Build age does not predict repair outcome.}
We compared older and newer cases within \ageComparedProjects{} projects.
Older cases had a higher unresolved rate in \ageOlderHigherProjects{}
projects and a lower rate in \ageOlderLowerProjects{}.
The rates were equal in the remaining \ageTiedProjects{} projects.
The mean difference in unresolved rates between the older and newer
halves was \ageMeanDifference{} percentage points. The exact sign test
gave $p=\ageSignTestP{}$, providing no evidence that older cases were harder to repair within the same project.

\textbf{Repair outcomes vary substantially across projects.} Among the
\projectComparedCount{} projects with at least twenty attempted cases, the
unresolved rate ranged from \projectMinUnresolvedRate{}\% for \code{harfbuzz}
to \projectMaxUnresolvedRate{}\% for \code{envoy},
\code{jackson-dataformats-binary}, \code{jackson-dataformats-text}, and
\code{suricata}.

\begin{rqanswer}{RQ1}
Historical reconstruction failed for \failRate{}\% of the \totalBuilds{} cases
we executed. Failures were highly concentrated: groups containing at least five
cases covered \concShare{}\% of the failure-analysis dataset. Re-execution
without a repair successfully built \noPatchRate{}\% of attempted cases. We found no consistent relationship between build age and repair outcome, but unresolved
rates varied substantially across projects.
\end{rqanswer}
\clearpage

\section{RQ2: \rqtwo}
\label{sec:rq2}
\subsection{Motivation}
Reusable rules are useful only if they repair recurring failures and can be
derived with limited human effort. RQ2 examines how many builds these rules
restore, whether rule derivation can run without human intervention, and
whether the applied repairs were needed for recovery.

\subsection{Approach}
We evaluated two consecutive phases of rule derivation. The first phase lasted
approximately three months. One of the authors selected failure groups,
reviewed the evidence and rules produced by the LLM, and revised the proposals
when needed. During the second phase, \sys{} selected groups, prepared the
evidence, generated candidate rules, and validated them without human
intervention. This phase started with the registry produced during the
human-guided work. We ran this phase until it reached approximately the same number of rule-associated recoveries as the human-guided phase. This took two weeks.

For each failure group selected during the automated phase, \sys{} prepared
evidence from two or three representative cases and asked GPT-5.6 Terra to
propose a rule. The proposal specified both the conditions under which the rule
should apply and the change it should make. \sys{} checked the repair scope, applied the candidate rule, and rebuilt the representative cases. If the rule restored a build or exposed a different failure, the framework added the rule to the registry and applied it to other matching cases. \sys{} recorded when each rule entered the registry and which successful repair chains used the rule.

We report the human-guided and automated phases separately. We count a
recovery as \emph{rule-associated} when the successful execution used at least one repair rule. We count cases that built successfully without a repair as
\emph{no-patch recoveries}. For the automated phase, we also report recovery rates by ecosystem. We calculate each rate by dividing
rule-associated recoveries by the number of cases attempted in that
ecosystem during the automated phase.

The two phases took place at different times, attempted different cases,
and shared an expanding rule registry. The human-guided phase also
included development of the framework. We report how long each phase
took to reach a similar number of recoveries, but these differences
prevent a direct comparison of their speed.

Early in the campaign, we prioritized failure groups that appeared likely to
produce reusable rules. We consequently explored some ecosystems more
than others. We later focused on underrepresented ecosystems, primarily
those outside C and C++, until we had examined approximately half of the
failure groups in each major ecosystem. Within these ecosystems, we
continued to choose groups based on their repair potential.

Some cases built successfully on a retry without a repair, so we checked
whether the rules were needed for recovery. First, we identified cases
that had failed without a repair before succeeding with one. We then
conducted a paired control using cases with a saved repair chain from
the successful execution. We selected \controlEvaluatedPairs{} of these cases using a fixed seed and stratified the sample by language and failure class. We executed
each case once with the saved repair and once without it. We used separate data
directories and alternated which version ran first.

Finally, we examined whether repaired builds preserved the behavior needed by a
vulnerability benchmark. A successful build does not guarantee that the
recorded input will reproduce the historical crash. ARVO encountered the same
problem, including cases where the input did not trigger the crash on an
archived vulnerable binary~\cite{arvo}. We therefore compared crash-reproduction rates
between rule-associated and no-patch recoveries to
check whether applying rules made the historical crash less likely to
reproduce. We report the difference between these rates with its 95\%
confidence interval. We also report how many recoveries met the
\emph{benchmark-ready} criteria defined in Section~\ref{subsec:outcomes}.

\subsection{Results}
\textbf{Reusable rules restored hundreds of builds.} Across both phases, \productiveRules{} repair rules contributed to
\ruleRecoveries{} rule-associated recoveries.
Table~\ref{tab:phases} reports \humanRecoveries{} from the human-guided
phase and \autoRecoveries{} from the automated phase. The automated phase
reached a similar number of recoveries in two weeks without human
intervention.

Most successful repair chains were short.
\singleRuleChainRecoveries{} recoveries used one rule. The remaining
\multiRuleChainRecoveries{} used multiple rules, with a median chain length of
\multiRuleChainMedian{} rules and a mean of \multiRuleChainMean{}.

\begin{table}[t]
\caption{Distinct cases recovered during the human-guided and automated
campaign phases. ``With rules'' counts cases whose successful execution
applied at least one repair rule. ``No patch'' counts cases that succeeded
without a repair.}
\label{tab:phases}
\centering\footnotesize
\begin{tabular}{lrrr}
\toprule
Phase & With rules & No patch & Total \\
\midrule
Human-guided (about 3 months) & 231 & 202 & 433 \\
Automated (2 weeks) & 238 & 84 & 322 \\
\midrule
Total & 469 & 286 & 755 \\
\bottomrule
\end{tabular}
\end{table}

\textbf{The automated phase derived new repair rules.} Of the
\autoRecoveries{} rule-associated recoveries during the automated phase, \autoOnlyRecoveries{} used only newly derived rules produced by
the automated loop, \autoExistingOnlyRecoveries{} used only rules from the existing registry, and
\autoMixedRuleRecoveries{} used both. In total, \autoDerivedRules{} newly derived rules contributed to succesful repairs. The automated phase therefore benefited from the existing registry while also extending it.

\textbf{Recovery differed across ecosystems.}
Table~\ref{tab:ecosystems} reports rule-associated recovery during the
automated phase. C had the highest observed rate at \autoCRecoveryRate\%, followed by JVM at
\autoJVMRecoveryRate\% and C++ at \autoCppRecoveryRate\%. Rust had the lowest rate at \autoRustRecoveryRate{}\%.

\begin{table}[t]
\caption{Cases recovered with repair rules during the automated phase, by
ecosystem.}
\label{tab:ecosystems}
\centering\footnotesize
\begin{tabular}{lrrl}
\toprule
Ecosystem & Recovered/attempted & Rate & 95\% CI \\
\midrule
C & 19/66 & 28.8\% & [19.3, 40.6] \\
JVM & 50/212 & 23.6\% & [18.4, 29.7] \\
C++ & 119/692 & 17.2\% & [14.6, 20.2] \\
Go & 41/349 & 11.7\% & [8.8, 15.5] \\
Python & 2/49 & 4.1\% & [1.1, 13.7] \\
Rust & 7/309 & 2.3\% & [1.1, 4.6] \\
\bottomrule
\end{tabular}
\end{table}

\textbf{The paired control confirmed that some builds needed the repair.}
We identified \strictControlledRecoveries{} cases that failed without a
repair before succeeding with a repair chain. The paired control evaluated
\controlEvaluatedPairs{} pairs. One pair encountered an infrastructure
failure, leaving \controlComparableCases{} pairs with usable results
both with and without the repair. Among the \controlComparableCases{} comparable pairs, \controlRepairOnlyCases{} cases built successfully only with the repair.
The remaining \controlBothRecoveredCases{} succeeded both with and without it. No case succeeded only without the repair. The repair was required in \controlRepairRequiredRate\% of the comparable
pairs (95\% Wilson CI [\controlRepairRequiredCILow, \controlRepairRequiredCIHigh]). An exact McNemar test gave $p=\controlMcNemarP{}$.

We examined the final repair operation before each successful build. This
information was available for \mechanismClassifiedRecoveries{} of the \ruleRecoveries{} rule-associated
recoveries. In \mechanismEnvironmentChanges{} cases, the final operation changed
a dependency source, package version, base image, toolchain or another part of the build environment. In \mechanismRefreshRecoveries{} cases, it
retried an operation or refreshed a package index or cache. The remaining
\mechanismUnclassifiedRecoveries{} cases lacked enough detail to
classify the final operation.

\textbf{Repaired builds reproduced historical crashes at a similar 
rate to builds that succeeded without a repair.} Table~\ref{tab:fidelity} reports crash
reproduction and benchmark-ready outcomes. The recorded input reproduced the crash in \fidelityRepaired{}\% of rule-associated recoveries and
\fidelityNoPatch{}\% of no-patch recoveries. The difference was \fidelityDifference{} percentage
points, with a 95\% confidence interval of $[\fidelityDifferenceCILow,\fidelityDifferenceCIHigh]$. We therefore found no evidence that applying the repair rules reduced crash reproduction.

\begin{table}[t]
\caption{Historical vulnerability outcomes among recovered builds.
Benchmark-ready cases are a subset of cases reproducing the crash.}
\label{tab:fidelity}
\centering\footnotesize
\begin{tabular}{lrrr}
\toprule
Outcome Group & Recovered cases & Crash reproduced & Benchmark-ready \\
\midrule
No patch & 286 & 145 (50.7\%) & 127 (44.4\%) \\
Repair chain & 469 & 247 (52.7\%) & 205 (43.7\%) \\
\bottomrule
\end{tabular}
\end{table}

\begin{rqanswer}{RQ2}
Across both campaigns, \productiveRules{} repair rules contributed to
\ruleRecoveries{} restored builds. The automated campaign produced
\autoRecoveries{} recoveries in two weeks without human intervention,
compared with \humanRecoveries{} during approximately three months of
human-guided work. Of the automated campaign's recoveries,
\autoOnlyRecoveries{} used only newly derived rules. In the paired
control, \controlRepairOnlyCases{} of the \controlComparableCases{}
comparable cases needed the saved repair chain to build successfully.
Builds restored with rules reproduced their historical crashes at a
similar rate to builds that succeeded without a repair
(\fidelityRepaired{}\% versus \fidelityNoPatch{}\%).
\end{rqanswer}

\section{RQ3: \rqthree}
\label{sec:rq3}
\subsection{Motivation}
Per-case agents can inspect one repository and adapt its environment to the
failures they encounter~\cite{repo2run2025,executionagent2025}. This raises the question of why reusable rules are needed at all. RQ3 compares rules
and agents on the same failing cases and examines whether combining them
improves recovery and reduces repair effort. We measure build executions,
model usage, and which cases each approach restores. We also examine
how widely the rules are reused and whether they repair cases outside
those used during development.

\subsection{Approach}
We first measured how often the same repair rule helped restore different
cases, projects, and failure groups. Because a successful repair chain could contain more than one rule, we reported both recovered cases and \emph{rule credits}. We gave a
rule one credit each time it appeared in a successful repair chain. The number
of credits can therefore exceed the number of recovered cases. We also report how many projects and baseline failure groups used each rule in a successful repair.

We reserved \reserveCases{} previously unattempted cases to test whether the
registry could repair cases outside those used during rule development. We sampled by language and failure class using a fixed seed. We excluded Swift
because too few cases were available. We kept these cases out of rule derivation, validation, and subsequent repair campaigns.

After freezing the registry, \sys{} rebuilt each reserved case and
checked whether an existing rule matched the observed failure. These cases came from those that remained untouched after the campaign had examined many promising failure groups. This let us test the registry on the remaining failure families rather than on additional cases from groups where the rules had already succeeded.

Next, we compared the rules with per-case agents using cases that rules had restored during either campaign phase. We excluded the held-out sample and the RQ4 cases, leaving \pairedEligibleCases{} candidates whose recorded failures matched a rule in the current registry. We grouped these cases by language, failure class, and
failure-group size. We randomly ordered the cases within each group using a
fixed seed and sampled from the groups in turn.

\sys{} rebuilt candidates without a repair in the sampling order. We retained the first \pairedCases{} cases that still failed. This ensured that every case
presented an active repair problem when the comparison began. We used these initial executions to select the cases and excluded them from the reported build counts.

We ran the rule registry, Qwen3.8-27B, and GPT-5.6 Terra once on each retained
case. Qwen3.8 and Terra operated as per-case agents. \sys{} selected rules using the failure observed during the new execution. We ran the rule evaluation within one day of the agent runs to reduce changes in the external environment. We report recovered cases and build executions per recovery. For agents, we
also report model calls, tokens, and hosted-model charges. We report the
earlier cost of deriving the registry separately.

We report 95\% Wilson confidence intervals for the recovery rates.
Because we evaluated all three approaches on the same cases, we compared their
recovery outcomes using two-sided exact McNemar tests. We adjusted the three
pairwise \(p\)-values using the Holm-Bonferroni method.

Finally, we estimated the result of running the rules first and sending only the
remaining cases to an agent. For each case the rules did not restore, we used
the agent result and resource usage already observed for that case. Each agent started from the initial failure and did not inherit edits made by the rules. RQ4 evaluates agent escalation after deterministic repair stops.

\subsection{Results}
\textbf{A single derivation can serve many cases.}
Of the \appliedRuleCount{} rules applied during the campaigns, \productiveRules{} contributed to successful repair chains that restored \ruleRecoveries{} cases. The \topReusedRuleCount{} most frequently reused rules
accounted for \topReusedRuleCreditShare{}\% of all rule credits.
\crossProjectRuleCount{} rules helped restore builds across at least three projects and
accounted for \crossProjectRuleCreditShare{}\% of the credits. For example, a
rule that redirects unavailable GDAL netCDF archives contributed to
\gdalMirrorRecoveries{} recoveries. Once \sys{} validates a rule, it can apply that rule to other matching cases without further model calls.

\textbf{The registry did not restore the held-out cases.}
Table~\ref{tab:heldout} reports the results for the \reserveCases{} held-out cases. For \reserveNoMatchCases{} cases, no rule
matched the observed failure. We excluded a further
\reserveLibreOfficeCases{} LibreOffice cases because LibreOffice builds
repeatedly exceeded the execution timeout in earlier runs. These results show that repairing the remaining cases requires additional rules or per-case agent repair.

\begin{table}[t]
\caption{Outcomes when the frozen rule registry is applied to
\reserveCases{} previously unattempted cases.}
\label{tab:heldout}
\centering\footnotesize
\begin{tabular}{lr}
\toprule
Outcome & Cases \\
\midrule
No rule matched the observed failure & 376 \\
All matching rules tried; build unresolved & 8 \\
Rule matched; repair could not be applied & 5 \\
LibreOffice cases excluded from evaluation & 11 \\
\midrule
Total & 400 \\
Recovered & 0 \\
\bottomrule
\end{tabular}
\end{table}

The automated campaign spent \$\ruleDerivationCost{} on model calls
to derive rules, including proposals that did not enter the registry.
Dividing this cost by the campaign's \autoRecoveries{}
rule-associated recoveries gives approximately
\$\costPerAuto{} per recovery. Applying validated rules to further
cases requires no additional model calls.

\textbf{Rules restored similar numbers of builds with fewer executions.}
Table~\ref{tab:paired} compares the three approaches on the
\pairedCases{} failing cases. The rules restored \pairedRuleRecovered{} cases,
compared with \pairedQwenRecovered{} for Qwen3.8 and
\pairedTerraRecovered{} for GPT-5.6 Terra. Their recovery rates were
\pairedRuleRate\%, \pairedQwenRate\%, and \pairedTerraRate\%, respectively.
The corresponding 95\% Wilson confidence intervals were
[\pairedRuleCILow{}, \pairedRuleCIHigh{}] for the rules,
[\pairedQwenCILow{}, \pairedQwenCIHigh{}] for Qwen3.8, and
[\pairedTerraCILow{}, \pairedTerraCIHigh{}] for Terra.
None of the pairwise differences was statistically significant after the
Holm-Bonferroni adjustment (adjusted \(p \geq
\pairedMinAdjustedP{}\)).

The rules required \pairedRuleBuildRate{} builds per restored case, compared with \pairedQwenBuildRate{} for Qwen3.8 and \pairedTerraBuildRate{} for Terra. The agents therefore used \agentBuildLo{} to
\agentBuildHi{} times as many builds per recovery. Qwen3.8 made
\pairedQwenCalls{} model calls and used \pairedQwenTokens{} tokens. Terra
made \pairedTerraCalls{} calls, used \pairedTerraTokens{} tokens, and cost
\$\terraCostTotal{}. After excluding \pairedQwenHarnessFailures{} cases affected by Qwen3.8 harness failures, we compared the remaining
\pairedValidCases{} cases across all three approaches. The rules restored
\pairedValidRuleRecovered{}, Qwen3.8 restored
\pairedValidQwenRecovered{}, and Terra restored
\pairedValidTerraRecovered{}.

\begin{table}[t]
\caption{Rules, per-case agents, and estimated rules-first combinations on the
\pairedCases{} cases that still fail at the start of the experiment. For each
combination, we add the rule execution to the recorded agent work on the
\pairedRuleUnrecovered{} cases not restored by rules. Build counts include
attempts that did not restore the case. Model costs exclude the earlier cost of
deriving the registry.}
\label{tab:paired}
\centering\footnotesize
\setlength{\tabcolsep}{4pt}
\begin{tabular}{lrrrrr}
\toprule
 & Rules & Qwen3.8 & Terra & Rules, then Qwen3.8 & Rules, then Terra \\
\midrule
Restored / \pairedCases{} & \pairedRuleRecovered{} & \pairedQwenRecovered{} & \pairedTerraRecovered{} & \hybridQwenRecovered{} & \hybridTerraRecovered{} \\
Builds & \pairedRuleBuilds{} & \pairedQwenBuilds{} & \pairedTerraBuilds{} & \hybridQwenBuilds{} & \hybridTerraBuilds{} \\
Builds per restored case & \pairedRuleBuildRate{} & \pairedQwenBuildRate{} & \pairedTerraBuildRate{} & \hybridQwenBuildRate{} & \hybridTerraBuildRate{} \\
Agent tokens & 0 & \pairedQwenTokens{} & \pairedTerraTokens{} & \hybridQwenTokens{} & \hybridTerraTokens{} \\
Model API charge & \$0 & local & \$\terraCostTotal{} & local & \$\hybridTerraCost{} \\
\bottomrule
\end{tabular}
\end{table}

\textbf{Rules and agents restored different cases.}
The rules restored \pairedRuleOnlyVsQwen{} cases that Qwen3.8 missed and
\pairedRuleOnlyVsTerra{} that Terra missed.
Conversely, at least one agent restored \pairedAgentOnly{} cases that the rules
missed. Across all three approaches, \pairedAnyRecovered{} of the
\pairedCases{} cases were restored by at least one of them. Agent escalation
therefore added coverage.

\textbf{Combining rules and agents improved recovery and reduced agent work.}
We combined the recorded rule results with Qwen3.8's results on the cases the rules did not restore. This gave \hybridQwenRecovered{} recoveries using \hybridQwenBuilds{} builds, compared with \pairedQwenRecovered{} recoveries and \pairedQwenBuilds{} builds for Qwen3.8 alone. Sending only the unresolved cases to Qwen3.8 reduced token use from \pairedQwenTokens{} to \hybridQwenTokens{}.

Combining the recorded rule results with Terra's results on unresolved cases gave \hybridTerraRecovered{} recoveries using \hybridTerraBuilds{} builds. Terra alone restored \pairedTerraRecovered{} cases using \pairedTerraBuilds{} builds. The combination reduced token use from \pairedTerraTokens{} to
\hybridTerraTokens{} and API charges from \$\terraCostTotal{}
to \$\hybridTerraCost{}.

\begin{rqanswer}{RQ3}
Rules and per-case agents restored similar numbers of builds on the
\pairedCases{} failing cases, but agents required \agentBuildLo{}
to \agentBuildHi{} times as many build executions per recovery. The approaches restored different cases: at least one agent restored
\pairedAgentOnly{} cases missed by the rules. Combining the recorded outcomes gave \hybridTerraRecovered{} to
\hybridQwenRecovered{} recoveries while using fewer builds and less
than half the agent tokens of an agent alone. The registry restored
none of the \reserveCases{} held-out cases, showing the need to
derive additional rules or use per-case agents for the remaining
failures.
\end{rqanswer}
\clearpage

\section{RQ4: \rqfour}
\label{sec:rq4}
\subsection{Motivation}
RQ3 showed that rules and agents restore different cases and that using the
rules first can reduce agent work. \sys{} escalates unresolved cases after deterministic repair stops. RQ4 examines the cases that reach this point. We
first identify why rule-based repair stops and whether per-case agents
can repair the unresolved cases.

\subsection{Approach}
\label{sec:agents}
We first examined cases where a rule matched the failure but could not
apply its edit to the project's build files. We counted these application
failures across all recorded attempts and manually inspected the logs
and repair specifications for \applicationAuditCases{} affected cases
to identify why the edits failed.

We also examined whether the derivation loop repeatedly proposed
variations of the same repair without restoring the build. We identified
\repeatedRepairConcepts{} recurring strategies: Rust nightly and
\code{RUSTC\_BOOTSTRAP} handling, Go toolchain installation, retired Go
base builders, retired Rust base builders, and Kubernetes Go toolchain
handling. For each strategy, we compared how often its rules exposed
a different failure with how often they restored a build.

We selected \escalationCases{} unresolved cases to test whether agents
could repair them after deterministic repair stopped. The sample included
\methodExhaustedCases{} \emph{rule-exhausted} cases, where the loop
had tried every matching rule, and \scopeLimitedCases{}
\emph{scope-limited} cases, where the evidence suggested that a repair
required a case-specific choice our general rules could not express. We excluded cases blocked by unavailable external resources or infrastructure failures, since these prevented the agent from testing a repair.

Before running the agents, \sys{} rebuilt all \escalationCases{} cases
without a repair. We excluded \escalationProbeRecovered{} cases that
built successfully and evaluated the three main models on the remaining
\escalationEvaluatedCases{} failing cases. We ran Qwen3.8-27B, Qwen3.6-27B, and GPT-5.6 Terra on the same
\escalationEvaluatedCases{} cases. The two Qwen models have the same
size, allowing us to compare model generations with the same settings.
We also tested Terra, which had proposed the repair rules, as a per-case
agent. To examine whether a more capable model could repair additional
cases, we ran GPT-5.6 Sol on a diagnostic subset of \solSubsetCases{}
cases. We selected \solRecoveredCases{} cases that Qwen3.8 had restored,
\solEarlyStopCases{} where it had stopped early, and
\solProgressedCases{} where it had changed the failure without
restoring the build. We report 95\% Wilson confidence intervals for the recovery rates.
Because we evaluated the three main models on the same cases, we compared their
recovery outcomes using two-sided exact McNemar tests. We adjusted the three
pairwise \(p\)-values using the Holm-Bonferroni method.

\subsection{Results}
\textbf{Rules sometimes matched a failure but could not apply the repair.}
The applicator failed to make the requested edit in
\applicationFailureCount{} of \repairAttemptCount{} recorded attempts
(\applicationFailureRate{}\%). These failures affected
\applicationFailureCases{} cases, of which
\applicationUnresolvedCases{} remained unresolved. We manually inspected
\applicationAuditCases{} affected cases. In
\applicationAnchorFailures{} cases, the applicator could not find the
text or insertion point specified by the rule. These included
\regexAnchorFailures{} replacements, \beforeAnchorFailures{}
insertions before a missing line, and \afterAnchorFailures{}
insertions after a missing line. Supporting more build-file layouts could
therefore allow the applicator to attempt these repairs on more cases.

\textbf{The derivation loop sometimes repeated repairs without restoring a build.}
The registry contained \repeatedRuleIdentifiers{} rule identifiers for
the \repeatedRepairConcepts{} strategies described above. For three of these
strategies, \repeatedNoRecoveryIdentifiers{} identifiers exposed a
different failure \repeatedNoRecoveryProgressions{} times without
restoring a build. Go toolchain installation and Rust nightly handling restored
\goToolchainRecoveredCases{} and \rustNightlyRecoveredCases{} cases, but only
\repeatedSuccessfulIdentifiers{} identifiers contributed to those recoveries. \sys{} limits further attempts when no eligible rule remains or successive attempts return the same failure.

\textbf{Agents restored additional cases after deterministic repair stopped.}
Table~\ref{tab:escalation} reports results on the
\escalationEvaluatedCases{} cases. Qwen3.8 restored
\rqfourQwenRecovered{}, Qwen3.6 restored
\rqfourQwenPreviousRecovered{}, and Terra restored
\rqfourTerraRecovered{}, giving recovery rates of
\rqfourQwenRate{}\%, \rqfourQwenPreviousRate{}\%, and
\rqfourTerraRate{}\%, respectively. The recorded input reproduced
the historical crash in \rqfourQwenCrashReproduced{},
\rqfourQwenPreviousCrashReproduced{}, and
\rqfourTerraCrashReproduced{} of those recoveries, respectively.

\begin{table}[t]
\caption{RQ4 results on the same \escalationEvaluatedCases{} cases unfixed by reusable rules. Token
totals include conversation history sent again on later turns. Terra's cost is
estimated from recorded API usage.}
\label{tab:escalation}
\centering\footnotesize
\setlength{\tabcolsep}{3pt}
\renewcommand{\arraystretch}{0.9}
\begin{tabular}{lrrr}
\toprule
Measure & Qwen3.8 & Qwen3.6 & Terra \\
\midrule
Recovered / \escalationEvaluatedCases{} & \rqfourQwenRecovered{} & \rqfourQwenPreviousRecovered{} & \rqfourTerraRecovered{} \\
Rate & \rqfourQwenRate\% & \rqfourQwenPreviousRate\% & \rqfourTerraRate\% \\
95\% Wilson CI & [18,41] & [15,37] & [10,29] \\
Crash reproduced & \rqfourQwenCrashReproduced{} & \rqfourQwenPreviousCrashReproduced{} & \rqfourTerraCrashReproduced{} \\
Build executions & \rqfourQwenBuilds{} & \rqfourQwenPreviousBuilds{} & \rqfourTerraBuilds{} \\
Builds per recovered case & \rqfourQwenBuildRate{} & \rqfourQwenPreviousBuildRate{} & \rqfourTerraBuildRate{} \\
Model calls & \rqfourQwenCalls{} & \rqfourQwenPreviousCalls{} & \rqfourTerraCalls{} \\
Total tokens & 16.50M & 24.31M & 23.17M \\
API usage cost & Local & Local & \$\rqfourTerraCost{} \\
\bottomrule
\end{tabular}
\end{table}

None of the differences between models was statistically significant after the
Holm-Bonferroni correction. The largest difference was between Qwen3.8 and
Terra: Qwen3.8 restored \rqfourQwenOnlyVsTerra{} cases that Terra did not,
while Terra restored none that Qwen3.8 missed. The raw \(p\)-value was
\rqfourQwenTerraRawP{}, and the adjusted \(p\)-value was
\rqfourQwenTerraAdjustedP{}. The other adjusted \(p\)-values were at least
\rqfourOtherAdjustedP{}. All three models restored \rqfourAllRecovered{} of the
same cases, and \rqfourUnionRecovered{} were restored by at least one model.

The execution harness did not produce a reliable outcome for
\rqfourHarnessFailures{} cases. After excluding the affected runs, we retained
\rqfourQwenValidCases{} usable Qwen3.8 results,
\rqfourQwenPreviousValidCases{} Qwen3.6 results, and
\rqfourTerraValidCases{} Terra results. None of the excluded runs was counted as a recovery, so the total number of recoveries remained unchanged.

Although the models received the same limits, they used their budgets
differently. The token limit stopped
\rqfourQwenTokenStops{} Qwen3.8 runs,
\rqfourQwenPreviousTokenStops{} Qwen3.6 runs, and
\rqfourTerraTokenStops{} Terra runs. Qwen3.8 required
\rqfourQwenBuildRate{} builds per recovery, Qwen3.6 required
\rqfourQwenPreviousBuildRate{}, and Terra required
\rqfourTerraBuildRate{}. Terra cost
\$\rqfourTerraCostPerRecovery{} per recovered case.

In some cases, the models' explanations for stopping were not supported by empirical evidence. We checked \rqfourDeclaredStops{} cases where a model stopped after reporting an infrastructure problem or an unavailable resource.
Another model restored \rqfourContradictedStops{} of these cases. We therefore treat these explanations as each model's assessment rather than verified causes.

\textbf{The flagship model did not improve recovery on the diagnostic subset.}
Of the \solSubsetCases{} selected cases, \solProbeRecovered{} built without a
repair when re-executed. This left \solEvaluatedCases{} cases for the
comparison. The provider rejected one case under its cybersecurity policy, so
Sol completed \solExecutedCases{} runs. Sol restored
\solFinalRecovered{} cases, compared with \solSubsetQwenRecovered{} for
Qwen3.8, \solSubsetQwenPreviousRecovered{} for Qwen3.6, and
\solSubsetTerraRecovered{} for Terra. None of the pairwise differences was
statistically significant, with all raw \(p\)-values being at least
\solPairwiseMinP{}. Every case restored by Sol was also restored by at least one other
model. Sol used \solFinalBuilds{} builds, \solCalls{} model calls, and
\solTokens{} tokens. Its API usage cost \$\solCost{}. The recorded input reproduced the historical crash in \solCrashReproduced{} of Sol's recoveries.

\begin{rqanswer}{RQ4}
We found that deterministic repair struggled to apply edits across
different build-file layouts and sometimes repeated unsuccessful repair
strategies. The applicator failed to make the requested edit in
\applicationFailureRate{}\% of recorded attempts. Agents restored
\rqfourTerraRecovered{} to \rqfourQwenRecovered{} of the
\escalationEvaluatedCases{} cases left unresolved by deterministic
repair, giving recovery rates of \rqfourTerraRate{}\% to
\rqfourQwenRate{}\%. These results show that agent escalation can
extend repair beyond the existing rules. Sol did not improve recovery
on the diagnostic subset.
\end{rqanswer}

\section{Discussion}
\label{sec:discussion}
\subsection{Why Not Use an Agent for Every Build?}
Rules and per-case agents both use model reasoning to decide how to repair
a build. A per-case agent repeats that reasoning for every build. \sys{} uses evidence from representative failures to derive a rule that specifies when a
repair applies and what to change. Once validated, the rule can be inspected and applied to
other matching cases without further model calls.

Our paired evaluation supports using rules before agents. The rules
restored similar numbers of cases with fewer build executions, while
agents restored additional cases that the rules missed. Combining
the recorded outcomes gave \hybridTerraRecovered{} to
\hybridQwenRecovered{} recoveries out of \pairedCases{} cases,
using fewer builds and less than half the agent tokens of an agent
alone. Rules handled recurring failures efficiently, while agent escalation
provided flexibility on the remaining cases.

Building the registry also takes work. Each new rule requires model
calls and validation builds, and some repairs require extending the
applicator with new editing operations. We paid these costs during
framework development and rule derivation. Reuse becomes worthwhile
when enough cases share a failure to justify that initial effort.
A larger registry can reduce repeated diagnosis, but it still needs
maintenance as build tools and external resources change.

The held-out evaluation showed where the registry needed further work:
no rule matched \reserveNoMatchCases{} of the \reserveCases{} cases.
Deriving another rule is most useful when several cases share a failure.
For a failure affecting only one case, an agent may repair the build
with less effort than deriving and validating a reusable rule.
These limits motivate the division of work in \sys{}: apply validated rules wherever they match, then escalate what remains.

\subsection{Improving the Repair Process}
Our application audit showed that rules sometimes recognized a failure
but could not find the text they needed to change. Parsers for package manifests and build instructions could help the applicator make the
same edit across different file layouts. The resulting repair would still
need to be validated through execution, since a more portable edit could also
spread an incorrect change more widely.

The derivation loop sometimes proposed several versions of the same
unsuccessful repair. Keeping a record of the strategies already tried
and their execution results could help the model choose a different
approach. The framework could also use this history to stop rule
derivation or send the case to an agent when repeated proposals
fail to restore the build.

Per-case agents remained useful after the registered rules stopped. They
restored \rqfourTerraRate{} to \rqfourQwenRate{}\% of the remaining cases.
However, more model reasoning cannot recover a missing external resource or
correct a failure in the execution infrastructure. The models reported an
infrastructure or resource problem in \rqfourDeclaredStops{} cases, but another
model restored \rqfourContradictedStops{} of them. The framework should therefore validate the model's explanation for stopping using the execution logs before deferring a case.

Improving the agent harness or using stronger models may increase
recovery. Many unsuccessful runs reached the token limit, so better
evidence selection, conversation management, and tool use could help
agents spend their budgets more effectively. Sol did not improve
recovery on our small diagnostic subset. Further experiments could
test harness changes with the same model and compare models using
the same harness.

\subsection{Preserving Executable Benchmarks}
Benchmark maintainers can reduce future repair work by preserving
the resources needed to rebuild each case. Alongside the source
revision, they should record base-image digests, dependency revisions,
toolchain versions, and build instructions. Where redistribution
permits, archiving installers and source packages would also protect
against downloads becoming unavailable.

Benchmark maintainers should also preserve build logs, repair chains,
and the results of running the recorded input at both revisions.
These records let researchers inspect earlier repairs and check
whether a restored build still reproduces the historical vulnerability
and meets the benchmark-ready criteria.

\section{Threats to Validity}
\label{sec:threats}
\subsection{Internal Validity}
\textbf{Attribution.} Some cases built successfully on retries without a repair. We therefore report no-patch recoveries separately and use a paired control to check whether repairs were needed. For this reason, we also describe recoveries as \emph{rule-associated} rather than claiming that every applied rule caused the observed recoveries.

\textbf{Campaign comparison.} The automated campaign started with the registry developed during
the human-guided phase. The phases attempted different cases at
different times, and the human-guided work also included framework
development. Early selection prioritized promising failure groups,
while later selection focused more on underrepresented ecosystems.
These differences limit conclusions about how much automation
reduced repair time.

\textbf{Changes between executions.} External resources continued
to change during the study. Some previously failing cases built successfully without a repair when we re-executed them before RQ3 and RQ4. We excluded these cases because they no longer presented a repair problem. In RQ3, we ran the rule and agent evaluations within one day of each other. These steps reduce, but do not eliminate, differences caused by changes in
package repositories, mirrors, and other external services.

\textbf{Harness errors.} We inspected failed runs to distinguish repair failures from errors in the experimental harness. We report results for all selected cases and additional comparisons excluding runs affected by harness
errors. We count patches that fail to apply as repair outcomes.
Some errors require judgment to classify, so we may have
misclassified cases despite inspecting their logs.

\subsection{Construct Validity}
\textbf{Failure and success measures.} We count a case as failed
when it does not complete the full reconstruction and verification process.
This definition includes build failures, unavailable resources, and failures
in the execution infrastructure. Some failures may be temporary or have
causes unrelated to environment decay. Conversely, a successful build does
not by itself restore a vulnerability benchmark. We therefore report crash reproduction and the stricter benchmark-ready outcome separately. These checks do not prove that every behavior of the rebuilt program is unchanged.

\textbf{Grouping and rule credit.} Normalized signatures can merge distinct
causes or separate variants of one cause. Shared groups motivate investigation
but do not guarantee a common repair. Rule credits can overlap within a chain,
so we distinguish them from distinct cases. Several rule identifiers may also describe the same repair strategy. Models may misidentify why they stopped, so we check their explanations against the execution evidence.

\textbf{Manual audits.} We classified repair operations and repair limits using the available logs and repair specifications. Some earlier records contained less detail, and different rule identifiers sometimes described similar
repair strategies. These classifications required judgment, and
another researcher might group some operations differently.

\textbf{Costs and budgets.} Build executions vary in duration and resource use. Some fail early, while others compile large projects. Token totals count conversation history again each time the harness sends it to the model, and
different tokenizers may count the same text differently. Reported
model charges exclude hardware and human effort. We report rule
derivation costs separately from the cost of applying existing rules.
Our cost comparisons therefore cover only part of the total effort
needed to develop and run the framework.

\subsection{External Validity and Repeatability}
\textbf{Dataset scope.} All cases use OSS-Fuzz build recipes and our extended ARVO workflow. Their shared base images and build conventions may make rules easier to reuse than in other build systems. We also deliberately increased the
representation of non-C/C++ projects. The \failRate{}\%
reconstruction-failure rate therefore describes our dataset and may
differ from the rate across the full OSS-Fuzz issue tracker.

\textbf{Failure and repair selection.} The failure-analysis
dataset contains cases for which \sys{} produced a normalized observation. We assembled it first by including failures from non-C/C++ projects and then
adding C/C++ cases. The repair campaigns prioritized recurring groups
that appeared likely to support reusable rules. These choices affect the failure groups and ecosystem recovery rates we report.
We selected the held-out cases from those the campaigns had not
attempted. This limits how well the evaluation represents rule reuse
across the wider dataset.

\textbf{Evaluation cohorts.} RQ3 used cases that had previously
been repaired by rules and still failed when re-executed. Most selected cases were C or C++ builds, and selecting previously repaired cases may favor the rule-based approach. RQ4 used cases where the rules had stopped or could not express the required repair. Recovery rates may differ on other
samples. We also did not compare \sys{} with reconstruction using
archived distribution snapshots.

\textbf{Model repeatability.} We ran each model once per case.
The local models used temperature 1.0, so repeated runs could produce different results. Both Qwen releases use the same serving settings, but
their weights and templates differ. Hosted models may also
change over time. We selected the small Sol subset using earlier Qwen outcomes, and many Sol runs reached the token limit. A larger evaluation is needed
to assess how model capability affects repair success. Models may
also have encountered the public projects and issue reports during
training, which could influence their repairs.

\section{Related Work}
\label{sec:related}
\subsection{Historical Build Reproducibility}

Prior work has studied how to preserve software builds and reproduce
their outputs. Reproducible-build research has examined whether
rebuilding the same source produces bit-for-bit identical
artifacts~\cite{lamb2022reproducible,fourne2023flossing,
malka2025functional}. Studies of container reproducibility have
shown that rebuilding can still depend on changing image tags,
package repositories, and external downloads~\cite{malka2026docker}.
Researchers have addressed environment preservation through
functional package managers~\cite{malka2024reproducibility} and
developed systems to reconstruct dependencies for Python programs
and notebooks~\cite{horton2019dockerizeme,ye2022pyego,
wang2021snifferdog}. We examined how to restore historical builds
across several ecosystems after their environments had stopped
working.

Research on executable defect datasets has highlighted the need to
preserve historical build environments. Defects4J, BugSwarm, and
Vul4J provided historical failures for controlled
experimentation~\cite{just2014defects4j,tomassi2019bugswarm,
bui2022vul4j}, while Zhu and Rubio-Gonz\'alez showed that dataset
artifacts could lose reproducibility over
time~\cite{zhu2023reproducibility}. ARVO addressed this problem for
OSS-Fuzz vulnerabilities by reconstructing vulnerable and fixed
revisions~\cite{arvo}. CyberGym and SEC-bench subsequently used
executable historical software to evaluate security
agents~\cite{cybergym,secbench}. We extended ARVO to repair the
environment failures that remained after its reconstruction
mechanisms ran. This also distinguished our work from systems
such as CodeRover-S, which repaired vulnerabilities by changing
the target source~\cite{coderovers} whereas we preserved that source and
repaired the environment needed to build and execute it.

\subsection{Build Repair and Reusable Fixes}

Several studies have investigated build failures and their repair across
continuous integration, Docker, and other build
systems~\cite{rausch2017ci,vassallo2017ci,wu2020docker,
lou2020understanding}. A prior study by Nourry\etal also studied active OSS-Fuzz projects and identified failure causes across project source, build scripts, dependencies, toolchains, external resources, and execution
infrastructure~\cite{nourry2024}. These findings motivated us to
examine repairs across several stages of a build. We focused on
historical revisions, where changing the target source could alter
the benchmark case, and investigated whether environment repairs
could be reused across versions and projects.

Researchers have also developed tools to automate build repair using project history, dependency fixes, and reusable editing operations. HireBuild used
earlier changes to repair build scripts~\cite{hassan2018hirebuild},
while later work examined the limits of relying on history and
proposed repairs that did not require it~\cite{lou2019history}.
Dependency-focused systems included BuildMedic for Maven
builds~\cite{macho2018automatically} and PyDFix for Python dependency
errors~\cite{mukherjee2021pydfix}. Reusable operations have also
supported Dockerfile repair through human-guided
derivation~\cite{henkel2021shipwright} and mined
patterns~\cite{sakamoto2026dockerfile}. Other work has addressed
flaky Docker builds and equipped agents with operations for Android
build repair~\cite{flakidock2025,gradlefixer2026}. These approaches
have developed repair mechanisms for specific build technologies
and ecosystems. Alternatively, we examined how to derive and reuse repairs across the containers, dependencies, resources, toolchains, and scripts
involved in historical reconstruction.

Program-repair research has also investigated how to reuse fixes.
Refazer learned syntactic transformations from
examples~\cite{rolim2017refazer}, and Getafix learned fix patterns
from human-written changes~\cite{bader2019getafix}. These systems
showed how examples could guide repairs beyond the original case.
We applied this idea to build environments by asking an LLM to
derive rules from representative failures. We tested each candidate
by rebuilding cases, then reused accepted rules without further
model calls.

\subsection{Agents for Environment Setup and Compatibility Repair}

Recent studies have used language-model agents to automate repository
setup, compilation, and test execution~\cite{repo2run2025,
executionagent2025,milliken2025installamatic,hu2025compileagent,
yu2025cxxcrafter}. EnvBench has provided tasks for evaluating
environment setup in Python and JVM
projects~\cite{envbench2025}. Much of this work has focused on
preparing repositories for execution. RepoRescue examined a setting
closer to ours: older Python and Java repositories that no longer
worked in modern environments~\cite{reporescue2026}. It allowed
source changes and evaluated success through historical tests.
We preserved the target source revision and used the recorded
crashing input to check historical behavior. Other studies have
used agents or LLM-generated patches to repair dependency-related
build failures~\cite{macho2026agent,reyes2026byam}.

Recent work has explored how agents could reuse knowledge from
earlier setup tasks. SetupX transferred structured experience between
repositories, while BootstrapAgent recorded verified setup knowledge
for later agents~\cite{setupx2026,bootstrapagent2026}. We pursued
reuse through executable repair rules. Once a rule passed validation,
the framework could apply it to matching failures without another
model call. Per-case agents handled cases that remained unresolved.

These studies have advanced reusable repair and per-case environment
setup. We brought the two approaches together for historical build
restoration and evaluated where each helped. Our comparison examined
their recovery, build effort, model use, and the cases they restored.
Combining the recorded outcomes let us assess the benefit of
applying existing rules before spending agent effort on unresolved
cases.

\section{Conclusion}
\label{sec:conclusion}
We examined \totalBuilds{} historical OSS-Fuzz cases and found that
\failRate{}\% failed to complete reconstruction and verification
despite pinned source revisions. The failures were concentrated:
\concGroups{} groups covered \concShare{}\% of the failure-analysis
dataset. These recurring failures created opportunities to reuse
repairs across cases.

We developed \sys{} to derive reusable repair rules and escalate
unresolved cases to per-case agents. Across the campaigns,
\productiveRules{} rules contributed to \ruleRecoveries{} recoveries.
On the \pairedCases{} cases in our paired evaluation, agents restored
similar numbers of builds but required \agentBuildLo{} to
\agentBuildHi{} times as many build executions per recovery.
Combining the recorded outcomes gave \hybridTerraRecovered{} to
\hybridQwenRecovered{} recoveries with fewer builds and less than
half the agent tokens of an agent alone. Agents also restored
\rqfourTerraRate{} to \rqfourQwenRate{}\% of the cases left unresolved
by deterministic repair. These findings support applying reusable
rules first and using agents for the remaining cases.

Further work could improve how rules edit different build-file
layouts and how agents use their execution and token budgets.
Experiments could also vary the execution evidence given to
rule-authoring models and repair agents to determine which logs,
error excerpts, and repair history help them make better repairs.
Evaluating stronger models with the same harness would help identify
where model capability improves recovery. Throughout restoration,
checking the recorded historical behavior remains essential to
preserving the cases used by vulnerability benchmarks.

\section*{Data Availability}
We provide a partial replication package for review at
\url{https://zenodo.org/records/23491436}. It contains case manifests, normalized failure
observations, and data and analysis scripts for reproducing the main
quantitative results. The package documents which results the scripts
reproduce. We will release the full package upon publication, including the
complete rule registry, repair records, prompts, and model responses.

\begin{acks}
This work was supported by JSPS KAKENHI Grant Numbers JP24H00692, JP26H02500, JP25K03102, and JP26K02889.

The author used generative AI tools during the preparation of this paper. AI tools were used to generate the final versions of the figures from the author's sketches and data. AI tools were also used to assist with cross-checking reported numbers in the paper against the study data. Grammarly was used to check spelling and grammar. The author reviewed, edited, and verified all content and takes full responsibility for it.
\end{acks}

\bibliographystyle{ACM-Reference-Format}
\bibliography{veritas}
\end{document}